\documentclass[aps,showpacs,twocolumn,twoside,prc]{revtex4-1}

\usepackage{graphicx}
\usepackage{epstopdf}
\usepackage{amsmath}
\usepackage{amssymb}
\usepackage{bm}
\usepackage{xcolor,xspace}
\usepackage[ulem=normalem]{changes}
\usepackage{hyperref}

\usepackage{graphicx}
\newcommand{\ba}{\begin{eqnarray}}
\newcommand{\ea}{\end{eqnarray}}
\newcommand{\bsub}{\begin{subequations}}
\newcommand{\esub}{\end{subequations}}
\newcommand{\rs}[2]{\rho^2(E0; #1 \!\rightarrow\! #2) \cdot 10^3}

\def\ket#1{|#1\rangle}
\def\bra#1{\langle#1|}

\begin{document}
\title{Quadrupole Phonons in the Cadmium Isotopes}
\author{A.~Leviatan}\email{ami@phys.huji.ac.il}
\author{N.~Gavrielov}\email{noam.gavrielov@mail.huji.ac.il}
\affiliation{Racah Institute of Physics, The Hebrew University, 
Jerusalem 91904, Israel}

\author{J.E.~Garc\'\i a-Ramos}\email{enrique.ramos@dfaie.uhu.es}
\affiliation{Departamento de Ciencias Integradas y Centro de Estudios 
Avanzados en F\'isica, Matem\'atica y Computaci\'on, 
Universidad de Huelva, 21071 Huelva, Spain}

\affiliation{Instituto Carlos I de F\'{\i}sica Te\'orica y Computacional, 
Universidad de Granada, Fuentenueva s/n, 18071 Granada, Spain}

\author{P.~Van~Isacker}\email{isacker@ganil.fr}
\affiliation{Grand Acc\'el\'erateur National d'Ions Lourds,
  CEA/DRF-CNRS/IN2P3, Bvd Henri Becquerel, F-14076 Caen, France}

\date{\today}  

\begin{abstract}
A key question concerning the spherical-vibrator attributes of states 
in Cadmium isotopes is addressed by means of a boson Hamiltonian
encompassing U(5) partial dynamical symmetry. The U(5) symmetry is preserved
in a segment of the spectrum and is broken in particular non-yrast states,
and the resulting mixing with the intruder states is small.
The vibrational character is thus maintained in the majority of
low-lying normal states, as observed in $^{110}$Cd.
\end{abstract}

\maketitle

The concept of a phonon is indispensable
to understand collective behavior in quantum-mechanical many-body systems.
In particular in condensed matter, the description of lattice 
excitations requires the introduction of elementary modes of vibration
that are identified with phonons.
Phonons also play a central role in nuclear physics,
notably in the interpretation of the collective motion of nucleons in an 
atomic nucleus. 
A~standard model of the nucleus is in terms of a quantum liquid drop
that exhibits vibrations around an equilibrium shape,
which, if deformed, can also rotate.
In their seminal studies Bohr and Mottelson~\cite{Bohr52,Bohr53,Bohr75}
argued that the collective low-energy properties of nuclei
are dominated by quadrupole vibrations,
whose nature depends on the equilibrium shape.
Small oscillations about spherical equilibrium
can be described in terms of a single type of quadrupole phonon
while the oscillations about a quadrupole-deformed equilibrium
require the introduction of two different phonons
that generate so-called $\beta$ and $\gamma$ vibrations.

This Rapid Communication deals with vibrations of spherical nuclei.
The first observation to be made is that,
despite more than half a century of research,
the phonon interpretation of low-energy nuclear structure
remains controversial, as exemplified by the Cadmium isotopes. 
The latter since long have been considered
as archetypal examples of nuclei
that exhibit small-amplitude vibrations around a spherical shape,
to the extent that they have become textbook material
to illustrate nuclear phonon 
behavior~\cite{Bohr75,Iachello87,Bonatsos88,Casten00,Heyde04}. 
Evidence for near-harmonic vibrational properties of Cd isotopes
was reported for up to three~\cite{Aprahamian87}
and even up to six~\cite{Deleze93a} quadrupole phonons.
Nevertheless, it was also realized early on~\cite{Gneuss71}
that not all low-energy levels of these isotopes 
can be considered as vibrational
and that additional levels exist at low excitation energies.
The latter, named coexisting or intruder states~\cite{Heyde11},
were claimed to arise because of proton excitations across the $Z\!=\!50$
shell closure, character that was later confirmed in two-proton transfer 
reactions~\cite{Fielding77}. 
Over the years intruder bands were identified in many even-mass 
Cd isotopes~\cite{Kumpulainen92} and, in parallel, models were extended to 
include such states.
Extensive $E2$ decay patterns were established in several Cd isotopes
and reproduced theoretically, albeit laboriously, by allowing mixing
between vibrational and intruder states,
see e.g.\ Refs.~\cite{Heyde82,Deleze93b}.
However, as more data on the Cd isotopes were collected,
the interpretation in terms of vibration--intruder mixing became 
increasingly untenable:
decay properties of $^{112}$Cd could not be explained~\cite{Garrett07},
those of $^{114}$Cd were found to be ``enigmatic''~\cite{Casten92},
of $^{116}$Cd to be ``puzzling''~\cite{Kadi03}.
The crisis culminated in papers claiming the
``breakdown of vibrational motion''not only in the Cd~\cite{Garrett08}
but also in the neighboring Pd and Sn isotopes~\cite{Garrett10}.
This paradoxical behavior, characterized in~\cite{Heyde11} as an
unsolved problem, continues to attract considerable
attention~\cite{Garrett12,Batch12,Batch14,Papenbrock15,Garrett16,Prados17,
Nomura18}.
\begin{figure*}
\centering
\includegraphics[width=\linewidth,clip]{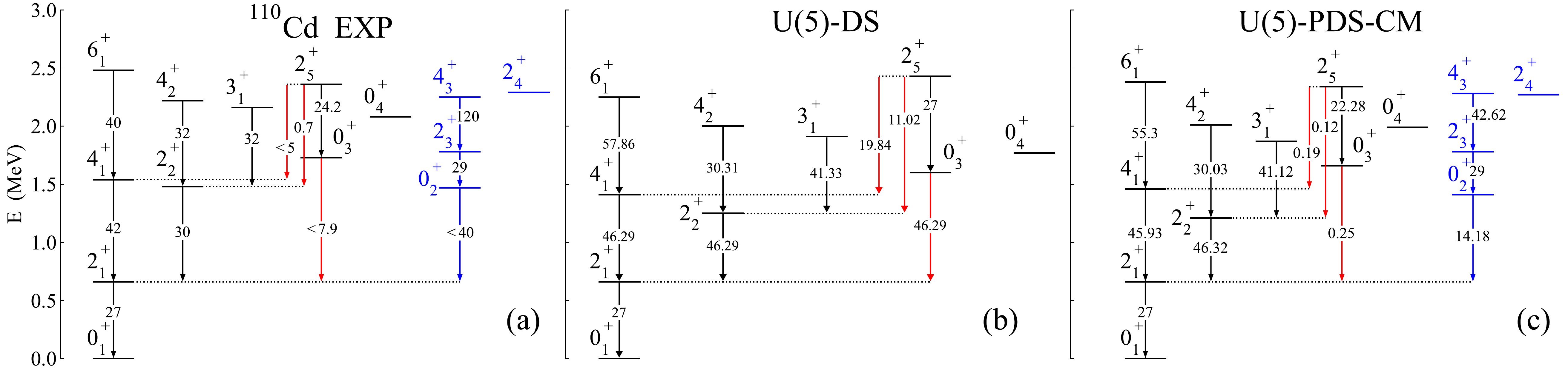}
\caption{\small
(a)~Experimental spectrum and representative 
$E2$ rates~\cite{Garrett12,NDS12} (in W.u.)  of normal 
and intruder levels ($0^{+}_2,\,2^{+}_3,\,4^{+}_3,\,2^{+}_4$)
in $^{110}$Cd. 
(b)~Calculated U(5)-DS spectrum obtained from 
$\hat{H}_{\rm DS}$~(\ref{H-DS}) with parameters 
$t_1\!=\!715.75,\, t_2\!=\!-t_3\!=\!42.10,\, 
t_4\!=\!11.38$ keV and $N\!=\!7$. 
(c)~Calculated U(5)-PDS-CM spectrum, obtained from 
$\hat{H}$~(\ref{Hfull}) with parameters 
$t_1\!=\!767.83,\, t_2\!=\!-t_3\!=\!73.62,\, 
t_4\!=\!18.47,\,r_0\!=\!2.15,\, e_0\!=\!-6.92,\,
\kappa\!=\!-72.73,\,\Delta\!=\!9978.86,\,\alpha\!=\!-42.78$ 
keV and $N\!=\!7\,(9)$ in the normal (intruder) sector. 
For a complete listing of $B(E2)$ values and choice 
of $E2$ parameters, see Tables~I-II.}
\label{fig-1}
\end{figure*}

In this Rapid Communication, we suggest that the vibrational
interpretation of the Cd isotopes can be resurrected not, as attempted 
previously, by mixing vibrational and intruder states but 
by mixing particular phonon states. 
From a formal point of view, 
the latter mechanism represents a departure from U(5),
which is the dynamical symmetry (DS) of spherical nuclei
in the collective model~\cite{Bohr75} and 
the interacting boson model (IBM)~\cite{Iachello87},
and generalizes it to a U(5) partial 
dynamical symmetry (PDS)~\cite{Leviatan11}.

A Hamiltonian with DS is written
as a linear combination of Casimir operators of nested algebras,
leading to complete solvability of its spectrum
with exact quantum numbers for all eigenstates~\cite{Iachello87,Iachello06}.
This property, although very appealing,
is rarely, if ever, satisfied in an existing quantum-mechanical system.
However, more realistic Hamiltonians can be constructed,
which satisfy the stringent DS conditions only partially.
This leads to three different types of PDS:
(i)~some eigenstates retain all quantum numbers~\cite{Alhassid92,Leviatan96},
(ii)~all eigenstates retain some quantum numbers~\cite{Leviatan86,Isacker99},
and (iii)~some eigenstates retain some quantum numbers~\cite{Leviatan02}.

In the following we apply a PDS of type (i)
to explain the spectroscopic properties of $^{110}$Cd.
The starting point is the U(5) limit of the IBM, 
corresponding to the following chain of nested 
algebras~\cite{Iachello87,Arima76},
\ba
{\rm U(6)\supset U(5)\supset SO(5)\supset SO(3)} ~. 
\label{U5-DS}
\ea
The basis states $\ket{[N],n_d,\tau,n_{\Delta},L}$ 
have quantum numbers which are the labels of irreducible 
representations of the algebras in the chain. 
Here $N$ is the total number of monopole ($s$) and quadrupole ($d$) 
bosons, $n_d$ and $\tau$ are the $d$-boson number and seniority, 
respectively, and $L$ is the angular momentum. 
The multiplicity label $n_{\Delta}$ counts the 
maximum number of $d$-boson triplets coupled to $L\!=\!0$. 
The U(5)-DS Hamiltonian has the form
\ba
\hat{H}_{\rm DS} = 
t_1\,\hat{n}_d + t_2\,\hat{n}_{d}^2
+ t_3\,\hat{C}_{{\rm SO(5)}} + t_4\,\hat{C}_{{\rm SO(3)}} ~,
\label{H-DS}
\ea
where $\hat{C}_{\rm G}$ denotes a Casimir operator of G, and 
$\hat{n}_d\!=\!\sum_{m}d^{\dag}_md_m\!=\!\hat{C}_{{\rm U(5)}}$. 
$\hat{H}_{\rm DS}$ is completely 
solvable for {\it any} choice of parameters $t_i$, with eigenstates
$\ket{[N],n_d,\tau,n_{\Delta},L}$ and eigen-energies
\ba
E_{\rm DS} =
t_1\, n_d + t_2\, n_{d}^2 
+ t_3\, \tau(\tau+3) + t_4\, L(L+1) ~.
\label{E-DS}
\ea
A typical U(5)-DS spectrum exhibits $n_d$-multiplets of 
a spherical vibrator, with a two-phonon ($n_d\!=\!2$) triplet of 
states ($L\!=\!4,2,0$) at an energy 
$E(n_d\!=\!2)\!\approx\! 2E(n_d\!=\!1)$ 
above the ground state ($n_d\!=\!L\!=\!0$) and first-excited state
($n_d\!=\!1,L\!=\!2$),
and a three-phonon ($n_d\!=\!3$) quintuplet of states 
($L\!=\!6,4,3,0,2$) at $E(n_d\!=\!3)\approx 3E(n_d\!=\!1)$.
A quadrupole operator proportional to
\ba
\hat{Q} = d^{\dag}s + s^{\dag}\tilde{d} ~,
\label{Q}
\ea
enforces strong ($n_d+1\!\rightarrow\! n_d$) $E2$ transitions 
with particular ratios, {\it e.g.}, 
$\frac{B(E2;\,n_d =2, L=0,2,4\rightarrow n_d=1,L=2)}
{B(E2;\,n_d=1,L=2\rightarrow n_d=0,L=0)}
=2\frac{(N-1)}{N}$.

The empirical spectrum of $^{110}$Cd, shown in Fig.~1(a), 
consists of both normal and intruder levels, the latter 
based on 2p-4h proton excitations across the $Z\!=\!50$ closed shell. 
Experimentally known $E2$ rates are listed in Tables~I and~II.
A comparison of the calculated spectrum [Fig.~(1b)] 
and $B(E2)$ values [Table~I], 
obtained from $\hat{H}_{\rm DS}$~(\ref{H-DS}),
demonstrates that most normal states have good spherical-vibrator
properties, and conform well with the properties of U(5)-DS. 
However, the measured 
rates for $E2$ decays from the non-yrast states, $0^{+}_3\,(n_d\!=\!2)$ 
and $[0^{+}_4,\, 2^{+}_5\,(n_d\!=\!3)]$, 
reveal marked deviations from this behavior. 
In particular, 
$B(E2;\,0^{+}_3\!\rightarrow\! 2^{+}_1) \!<\! 7.9$, 
$B(E2;\,2^{+}_5\!\rightarrow\! 4^{+}_1) \!<\! 5$,
$B(E2;\,2^{+}_5\!\rightarrow\! 2^{+}_2) \!=\! 0.7^{+0.5}_{-0.6}$ W.u.,  
are extremely small compared to the 
U(5)-DS values: $46.29$, $19.84$, $11.02$ W.u., respectively. 
Absolute $B(E2)$ values for transitions from the $0^{+}_4$ state 
are not known, but its branching ratio to $2^{+}_2$ is small.

Attempts to explain the above deviations in terms of 
mixing between the normal spherical [U(5)-like] states and 
intruder deformed [SO(6)-like] states have been shown to be 
unsatisfactory~\cite{Garrett08,Garrett12}. 
The reasons are two-fold. (i)~An adequate description of the 
two-phonon $0^{+}_3$ state requires strong (maximal $\sim$ 50\%) 
normal-intruder mixing 
which, in turn, results in serious disagreements with the observed 
decay pattern of three-phonon yrast states. 
(ii)~The discrepancy in the decays of the non-yrast two- 
and three-phonon states 
persists throughout the range $A\!=\!$ 110-126, 
including the heavier $^{A}$Cd isotopes~\cite{Batch12,Batch14}, 
even though the energy of intruder states rises away from 
neutron mid-shell, and the mixing is reduced. 
These observations have led to the conclusion that
the normal-intruder strong-mixing scenario needs to be rejected,
and have raised serious questions on the validity of the 
multi-phonon interpretation~\cite{Garrett08,Garrett12}. 
In what follows, we consider a possible explanation for the 
``Cd problem'', based on U(5)-PDS. 
The latter corresponds to a situation in which the U(5)-DS
is obeyed by only a subset of states and is broken in other
states. Similar PDS-based approaches
have been implemented in nuclear spectroscopy, in conjunction with the 
SU(3)-DS~\cite{Leviatan96,LevSin99,Casten14,levramisa13} and 
SO(6)-DS~\cite{Isacker99,Leviatan02,Ramos09} 
chains of the IBM. Here we focus on U(5)-PDS associated with 
the chain~(\ref{U5-DS}).
\begin{table}[t]
\begin{center}
\vspace{-1.1cm}
\caption{\label{TableI}
\small
Absolute (relative in square brackets) 
$B(E2)$ values in W.u.
for $E2$ transitions from normal levels in $^{110}$Cd.
The experimental (EXP) values are taken from~\cite{NDS12,Garrett12}. 
The U(5)-DS values are obtained for an $E2$ operator 
$e_{B}\,\hat{Q}$, Eq.~(\ref{Q}), with $e_{B}\!=\!1.964$ $(\rm W.u.)^{1/2}$.
The U(5)-PDS-CM values are obtained using $\hat{T}(E2)$, 
Eq.~(\ref{T-E2}), with $e_{B}^{(N)}\!=\!1.956$ and 
$e_{B}^{(N+2)}\!=\!1.195$ $(\rm W.u.)^{1/2}$. In both calculations the 
boson effective charges were fixed by the empirical 
$2^{+}_1\rightarrow 0^{+}_1$ rate. 
Intruder states $0^{+}_{2;i}\,2^{+}_{3;i},\,4^{+}_{3;i},\,2^{+}_{4;i}$,
are marked by a subscript~$i$.}
\begin{tabular}{lllcc}
\hline
$L_i$ $\;\;$ & $ L_f$ $\;\;$ & EXP $\;\;$ & U(5)-DS $\;$  & U(5)-PDS-CM \\[1pt]
\hline
$2^+_{1}$ & $0^+_{1}$ & 27.0 (8) & 27.00 & 27.00 \\[1pt]

$4^+_{1}$ & $2^+_{1}$ & 42 (9)   & 46.29 & 45.93 \\

$2^+_{2}$ & $2^+_{1}$ & 30 (5); 19 (4)\footnote{From Ref.~\cite{Garrett12}}
   & 46.29 & 46.32 \\
         & $0^+_{1}$ & 1.35 (20); 
0.68 (14)$^a$ & 0.00 & 0.00 \\
$0^+_{3}$ & $2^+_{2}$ & $<$ 1680$^a$  & 0.00 & 55.95 \\
         & $2^+_{1}$ & $<$ 7.9$^a$ & 46.29 & 0.25  \\
$6^+_{1}$ & $4^+_{1}$ & 40 (30); 62 (18)$^a$         & 57.86 & 55.30 \\
         & $4^+_{2}$ & $<$ 5$^a$ & 0.00  & 0.00 \\
         & $4^+_{3;i}$ & 14 (10); 36 (11)$^a$       &        & 2.39 \\         
$4^+_{2}$ & $4^+_{1}$ & 12$^{+4}_{-6}$; $^a$10.7$^{+4.9}_{-4.8}$  & 27.55 & 27.45 \\
         & $2^+_{2}$ & 32$^{+10}_{-14}$; 22 (10)$^a$  & 30.31 & 30.03 \\
	 & $2^+_{1}$ & 0.20$^{+0.06}_{-0.09}$; 0.14 (6)$^a$    & 0.00  & 0.00  \\
         & $2^+_{3;i}$ & $<$ 0.5$^a$ &        & 0.005 \\                
$3^+_{1}$ & $4^+_{1}$ & 5.9$^{+1.8}_{-4.6}$; $^a$2.4$^{+0.9}_{-0.8}$ 
& 16.53 & 16.48 \\
         & $2^+_{2}$ & 32$^{+8}_{-24}$; 22.7 (69)$^a$ & 41.33 & 41.12 \\
	 & $ 2^+_{1}$ & 1.1$^{+0.3}_{-0.8}$; 0.85 (25)$^a$   & 0.00  & 0.00 \\
                     & $ 2^+_{3;i}$ & $<$ 5$^a$  &        & 0.012 \\
$ 0^+_{4}$ & $2^+_{2}$ & [$<$ 0.65$^a$]  & 57.86 & 1.24 \\
                     & $2^+_{1}$ & [0.010$^a$]  & 0.00  & 31.76 \\
                     & $2^+_{3;i}$ & [100$^a$]  &        & 16.32 \\
$2^+_{5}$ & $0^+_{3}$ & 24.2 (22)$^a$  & 27.00 & 22.28 \\
                    & $4^+_{1}$ & $<$5$^a$  & 19.84 & 0.19 \\
                    & $2^+_{2}$ &$^a$0.7$^{+0.5}_{-0.6}$ & 11.02 & 0.12 \\
         & $2^+_{1}$ & 2.8$^{+0.6}_{-1.0}$                  & 0.00  & 0.00 \\
	 & $ 2^+_{3;i}$ & $<$ 5$^a$  &       & 0.002 \\
	 & 
$ 0^+_{2;i}$ & $<$ 1.9$^a$ &       & 0.20 \\[2pt]
\hline
\end{tabular}
\end{center}
\label{tab:transitions-normal}
\end{table}
\begin{table}[]
\begin{center}
\vspace{-1.1cm}
\caption{\label{TableII}
\small
$B(E2)$ values (in W.u.)
for $E2$ transitions from intruder levels in $^{110}$Cd.
Notation and relevant information on the observables shown, 
are as in Table~I.}
\begin{tabular}{lllc}
\hline
$L_i$ $\;\;$ & $ L_f$ $\;\;$ & EXP $\;\;$ & U(5)-PDS-CM \\[1pt]
\hline
$0^+_{2;i}$ & $2^+_{1}$ & 
$<$ 40\footnote{From Ref.~\cite{Garrett12}}     & 14.18 \\[1pt]
$ 2^+_{3;i}$ & $0^+_{2;i}$ & 29 (5)$^a$           & 29.00 \\
	    & $0^+_{1}$  & 0.31$^{+0.08}_{-0.12}$; 0.28 (4)$^a$  & 0.08 \\
            & $2^+_{1}$ & 0.7$^{+0.3}_{-0.4}$; $^a$0.32$^{+0.10}_{-0.14}$ & 0.00 \\
            & $2^+_{2}$ & $<$ 8$^a$              & 0.96 \\
$ 2^+_{4;i}$ & $2^+_{1}$ & 0.019$^{+0.020}_{-0.019}$ & 0.10 \\
$ 4^+_{3;i}$ & $2^+_{1}$ & 0.22$^{+0.09}_{-0.19}$; 0.14 (4)$^a$  & 0.49 \\
            & $2^+_{2}$ & 2.2$^{+1.4}_{-2.2}$; 1.2(4)$^a$ & 0.00 \\
            & $2^+_{3;i}$ & 120$^{+50}_{-110}$; 115 (35)$^a$   & 42.62 \\
& $4^+_{1}$   & 2.6$^{+1.6}_{-2.6}$; $^a$1.8$^{+1.0}_{-1.5}$
& 0.00 \\[2pt]
\hline
\end{tabular}
\end{center}
\label{tab:transitions-intruder}
\end{table}	

The lowest spherical-vibrator levels comprise three classes of states,
\bsub
\ba
{\rm Class\,A:}\quad && n_d=\tau=\!0,1,2,3\quad (n_{\Delta} = 0)~,
\label{ClassA}\\
{\rm Class\,B:}\quad && n_d = \tau+2=2,3\,\quad (n_{\Delta} = 0)~,
\label{ClassB}\\
{\rm Class~C:}\quad && n_d = \tau = 3\,\;\;\quad\qquad (n_{\Delta} = 1)~. 
\label{ClassC}
\ea
\label{ClassABC}
\esub
In the U(5)-DS calculation of Fig~1(b), applicable to normal states only, 
the ``problematic'' states 
$[0^{+}_3\,(n_d\!=\!2)$ and $2^{+}_5\,(n_d\!=\!3)]$  
belong to class~B, and $0^{+}_4\, (n_d\!=\!3)$ belongs to class~C.
The remaining ``good'' spherical-vibrator states 
$[0^{+}_1\,(n_d\!=\!0);\,2^{+}_1\,(n_d\!=\!1);\,
4^{+}_1,2^{+}_2\,(n_d\!=\!2);\,6^{+}_1,4^{+}_2,3^{+}_1\,(n_d\!=\!3)]$ 
belong to class~A. As mentioned, the spherical-vibrator interpretation 
is valid for most normal states in Fig.~1(a), but not all. 
We are thus confronted with a situation in which some states 
in the spectrum (assigned to class~A) obey the predictions of U(5)-DS, 
while other states (assigned to classes B and C) do not. 
These empirical findings signal the presence of a partial dynamical 
symmetry, U(5)-PDS.

The construction of Hamiltonians with U(5)-PDS follows the general 
algorithm~\cite{Ramos09,Alhassid92},
by identifying operators which annihilate particular sets of 
U(5) basis states.
In the present case, this leads to the following interaction:
\ba
\hat{V}_0 &=& r_0\,G^{\dag}_{0}G_{0}
+ e_{0}\,\left (G^{\dag}_0 K_0 + K^{\dag}_{0}G_0 \right ) ~,
\label{V0}
\ea
where $\textstyle{G^{\dag}_{0} \!=\! [(d^\dag d^\dag)^{(2)} d^\dag]^{(0)}}$, 
$K^{\dag}_{0} \!=\! s^{\dag}(d^{\dag} d^{\dag})^{(0)}$ 
and standard notation of angular momentum coupling is used. 
$\hat{V}_0$ of Eq.~(\ref{V0}) is in normal-ordered form and satisfies
\ba
\hat{V}_0\vert [N], n_d=\tau, \tau, n_{\Delta}=0, L \rangle = 0 ~,
\label{V0vanish}
\ea
with $L\!=\!\tau,\tau+1,\ldots,2\tau-2,2\tau$,
for any choice of parameters $r_0$ and $e_0$.
Equation~(\ref{V0vanish}) follows from the fact
that the indicated states have $n_d\!=\!\tau$ and $n_{\Delta}=0$, hence 
do not contain a pair or a triplet of $d$-bosons 
coupled to $L=0$ and, as such, are annihilated by 
$K_0$~\cite{Iachello87} and $G_0$~\cite{Talmi03}.

The states of Eq.~(\ref{V0vanish}), which include those of class~A,
form a subset of U(5) basis states, hence remain solvable eigenstates 
of the Hamiltonian
\ba
\hat{H}_{\rm PDS} &=& \hat{H}_{\rm DS} + \hat{V}_0 ~,
\label{H-PDS}
\ea
with good U(5) symmetry and energies given in 
Eq.~(\ref{E-DS}) with $n_d = \tau$. 
It should be noted that while $\hat{H}_{\rm DS}$~(\ref{H-DS}) is diagonal 
in the U(5)-DS chain~(\ref{U5-DS}), 
the $r_0$-term ($e_0$-term) in $\hat{V}_0$ connects states with 
$\Delta n_d\!=\!0$ and $\Delta\tau=0,\pm 2,\pm 4,\pm 6$ 
$(\Delta n_d\!=\!\pm 1$ and $\Delta\tau=\pm 1,\pm 3)$.
Accordingly, the remaining eigenstates of $\hat{H}_{\rm PDS}$~(\ref{H-PDS}), 
in particular those of classes B and C, 
are mixed with respect to U(5) and SO(5). 
The U(5)-DS is therefore preserved in a subset of eigenstates,
for any choice of parameters in $\hat{H}_{\rm PDS}$, but is broken in others.
By definition, $\hat{H}_{\rm PDS}$ exhibits U(5)-PDS. 
Cubic terms of the type present in $\hat{V}_0$, Eq.~(\ref{V0}), 
were previously encountered in IBM studies 
of triaxiality~\cite{heyde84,zamfir91}, 
signature splitting~\cite{levramisa13,Bonatsos88b}, 
band anharmonicity~\cite{Ramos09,Ramos00},
and shape-coexistence~\cite{LevDek16,LevGav17} in deformed nuclei. 
Such higher-order terms show up naturally in microscopic-inspired 
IBM Hamiltonians derived by a mapping from self-consistent mean-field 
calculations~\cite{Nomura12,Nomura13}. 

The effect of intruder levels can be studied in the framework of the 
interacting boson model with configuration mixing
(IBM-CM)~\cite{,DuvBar81,SambMoln82}.
The latter involves the space of normal states described by a 
system of $N$ bosons representing valence nucleon pairs,
and the space of intruder states described by a system of $N\!+\!2$ bosons, 
accounting for particle-hole shell-model excitations. 
This procedure has been used extensively in describing 
coexistence phenomena in 
nuclei~\cite{heyde95,Foisson03,ramos11,ramos14}. 
In the present study of $^{110}$Cd, 
the Hamiltonian in the normal sector is taken to be $\hat{H}_{\rm PDS}$ of 
Eq.~(\ref{H-PDS}), acting in a space of $N=7$ bosons. 
The Hamiltonian in the intruder sector is taken to be 
of SO(6)-type~\cite{heyde95},
\ba
\hat H_{\rm intrud} &=& \kappa \hat{Q}\cdot \hat{Q} + \Delta ~,
\label{H-intrud}
\ea
acting in a space of $N=9$ bosons, with $\hat{Q}$ given in Eq.~(\ref{Q}). 
A~mixing term between the $[N]$ and $[N\!+\!2]$ boson spaces is defined 
as~\cite{heyde95,Foisson03,ramos11,ramos14},
\ba
\hat V_{\rm mix} =  
\alpha \left [(s^{\dagger})^{2} + (d^{\dagger}d^{\dagger})^{(0)}\right ] 
+ {\rm H.c.} ~,
\label{Vmix}
\ea
where H.c. means Hermitian conjugate. The combined Hamiltonian for the 
two configurations has the form
\ba
\hat{H} = \hat{H}_{\rm PDS}^{(N)} + \hat{H}_{\rm intrud}^{(N+2)} 
+ \hat {V}_{\rm mix}^{(N,N+2)} ~.
\label{Hfull}
\ea
Here $\hat{\cal O}^{(N)} \!=\! \hat{P}^{\dag}_{N}\,\hat{\cal O}\,\hat{P}_{N}$ 
and $\hat{\cal O}^{(N,N')} \!=\! \hat{P}^{\dag}_{N}\,\hat{\cal O}\,\hat{P}_{N'}$ 
for an operator $\hat{\cal O}$, with $\hat{P}_{N}$ 
a~projection operator onto the $[N]$ boson space. 
In general, an eigenstate of $\hat{H}$, 
\ba
\ket{\Psi} = a\ket{\Psi_{n}^{(N)}} + b\ket{\Psi_{i}^{(N+2)}} ~,
\label{Psi}
\ea 
involves a mixture of normal ($n$) and 
intruder ($i$) components with $N$  and $N\!+\!2$ bosons, respectively. 
Similarly, the $E2$ operator is defined as
\ba
\hat{T}(E2) = e_{B}^{(N)}\,\hat{Q}^{(N)} + e_{B}^{(N+2)}\,\hat{Q}^{(N+2)} ~,
\label{T-E2}
\ea
with boson effective charges, $e_{B}^{(N)}$ and $e_{B}^{(N+2)}$.

The Hamiltonian of Eq.~(\ref{Hfull}) has nine parameters. However,
most of them only improve the fit to energies, 
but do not affect the structure of the states nor the calculated
$E2$ rates, which are the challenge in the Cd problem. The 6 parameters
($t_i,\, r_0,\, e_0$) of $\hat{H}_{\rm PDS}$~(\ref{H-PDS})
do not affect the U(5) purity
of class-A states and, for small $\alpha$, the
deviations from U(5)-DS in a few non-yrast states,
is governed solely by the $r_0$ and $e_0$ terms.
The comparison with the empirical data, discussed below,
constitutes a stringent test for these PDS-based terms.

The spectrum and $B(E2)$ values obtained with $\hat{H}$ (\ref{Hfull})
and $\hat{T}(E2)$ (\ref{T-E2}), are shown in Fig.~1(c) and Tables~I-II. 
As seen, the IBM-PDS-CM calculation provides a good description 
of the empirical data in $^{110}$Cd. 
The normal states of class~A retain 
good U(5) symmetry and quantum numbers, to a good approximation. 
Their $\ket{\Psi_n^{(N)}}$ part involves a single component 
with $n_d$ value as in Eq.~(\ref{ClassA}). 
The mixing with the intruder states is weak 
[small $b^2$ in Eq.~(\ref{Psi})] and increases with $L$. 
Specifically, $b^2\!=\!0.9,\,2.2,\,3.6,\,5.9,\,4.6,\,6.1\,\%$ 
for the $0^{+}_1,\,2^{+}_1,\,2^{+}_2,\,4^{+}_1,\,3^{+}_1,\,4^{+}_2$ states, 
respectively. 
The $6^{+}_1$ state experiences a somewhat larger mixing ($b^2\!=\!17.3\%$), 
consistent with its enhanced decay to the intruder $4^{+}_{3;i}$ state. 
The high degree of purity is reflected 
in the calculated $B(E2)$ values for transitions between class A states 
which, as seen in Table~I, are very similar to those 
of U(5)-DS. In contrast, the structure of the non-yrast states 
assigned originally to classes~B and C, whose decay properties show 
marked deviations from the U(5)-DS limit, changes dramatically.
Specifically, the $0^{+}_3$ and 
$0^{+}_4$ states, which in the U(5)-DS classification are members of the 
two-phonon triplet and three-phonon quintuplet, 
interchange their character, and the U(5) decomposition of their 
$\ket{\Psi_n^{(N)}}$ parts peaks at $n_d\!=\!3$ and $n_d\!=\!2$, respectively. 
Similarly, the $2^{+}_5$ and $2^{+}_6$ states, 
which in the U(5)-DS classification are members of the three-phonon 
quintuplet and four-phonon octet, 
interchange their character, and the U(5) decomposition of their 
$\ket{\Psi_n^{(N)}}$ parts peaks at $n_d\!=\!4$ and $n_d\!=\!3$, respectively.
The mixing with the intruder states is weak 
$(b^2\!=\!5.1\%,\,2.9\%,\,4.4\%)$ for the 
$(0^{+}_3,\,2^{+}_5,\,2^{+}_6)$ 
states, and somewhat larger $(b^2\!=\!18\%)$ for the $0^{+}_4$ state. 
The resulting calculated values:
$B(E2;\,0^{+}_3\!\rightarrow\! 2^{+}_1) \!=\! 0.25$, 
$B(E2;\,2^{+}_5\!\rightarrow\! 4^{+}_1) \!=\! 0.19$ and 
$B(E2;\,2^{+}_5\!\rightarrow\! 2^{+}_2) \!=\! 0.12$ W.u.,
are consistent with the measured upper limits:
$7.9,\,5$ and $0.7^{+0.5}_{-0.6}$ W.u., respectively.
The calculated decay $0^{+}_4\!\rightarrow\! 2^{+}_2$ is weaker than 
$0^{+}_4\!\rightarrow\! 2^{+}_1$, however, the observed dominant branching 
to the intruder $2^{+}_{3;i}$ state is not reproduced. This may indicate 
a different structure for the $0^{+}_4$ state ({\it e.g.}, a 4p-6h 
proton excitation as speculated in~\cite{Garrett12}).
The dominant $E2$ decays of the $2^{+}_6$ state (not shown in Fig.~1) 
are predicted to be 
$B(E2;\,2^{+}_6\!\rightarrow\! 0^{+}_4) \!=\! 24.28$, 
$B(E2;\,2^{+}_6\!\rightarrow\! 4^{+}_1) \!=\! 15.73$, 
$B(E2;\,2^{+}_6\!\rightarrow\! 2^{+}_2) \!=\! 9.27$ and 
$B(E2;\,2^{+}_6\!\rightarrow\! 4^{+}_{3;i}) \!=\! 3.56$ W.u.

A few monopole transition rates are experimentally known 
in $^{110}$Cd~\cite{Jig16}, expressed in terms of the quantity
$\rho(E0) \!=\! \tfrac{\bra{f}\hat{T}(E0)\ket{i}}{eR^2}$, 
with $R\!=\!1.2A^{1/3}$ fm. The corresponding $E0$ operator in the 
IBM-CM, can be transcribed~\cite{Zerguine12} in 
the form $\hat{T}(E0) = (e_{n}\, N' + e_{p}\, Z)\eta\,
(\hat{n}_{d}^{(N)} + \hat{n}_{d}^{(N+2)})$, 
where $N'$ ($Z$) are neutron (proton) numbers and $e_p\!=\!2e_n\!=\! e$.
The measured~\cite{Jig16} and calculated (in curly brackets) 
strengths are given by
$\rs{0^+_{2;i}}{0^+_1} \!<\! 31\, \{0.75\}$, 
$ \rs{2^+_{3;i}}{2^+_1} \!=\! 9(8)\, \{10\}$, 
$ \rs{4^+_{3;i}}{4^+_1} \!=\!  106^{+98}_{-91}\, \{36\}$.
The calculated strengths, obtained with $\eta=0.063$, reproduce
the measured values, within the quoted error bars.
The calculation predicts 
$\rho^2(E0)\cdot 10^3\!\sim\! 10$ 
for $0^{+}_{4}\!\to\! 0^{+}_{2;i}$ and $2^{+}_{4;i}\!\to\! 2^{+}_{2}$ 
$E0$ transitions, which have not been measured so far.

In summary, we have addressed a key question concerning the phonon 
structure of states in Cd isotopes.
Our results suggest that the vibrational interpretation 
of $^{110}$Cd can be salvaged by introducing
a boson Hamiltonian that mixes particular phonon states
while keeping the mixing with coexisting intruder levels weak.
The proposed scheme relies on a partial dynamical U(5) symmetry,
in which most low-lying normal states in $^{110}$Cd maintain their 
spherical-vibrational character and only a few specific
non-yrast states exhibit a departure from U(5), in line with the 
empirical data.
Work currently in progress appears to indicate that the
same PDS-based approach can be implemented also in other
neutron-rich Cd isotopes, at least as an appropriate 
starting point for further refinements.

This work is supported in part (A.L. and N.G.) by the Israel Science 
Foundation (Grant 586/16) and (J.E.G.R.) by the Spanish 
Ministerio de Econom\'{\i}a y Competitividad and the European regional 
development fund (FEDER) under Projects No. FIS2014-53448-C2-2-P and 
by Consejer\'{\i}a de Econom\'{\i}a, Innovaci\'on, 
Ciencia y Empleo de la Junta de Andaluc\'{\i}a (Spain) 
under Groups FQM-370.

\end{document}